\documentclass[preprint,showpacs,superscriptaddress]{revtex4}

\usepackage{graphicx}
\usepackage{dcolumn}
\usepackage{bm}

\begin{document}


\title{Topological phase transition of single-crystal Bi based on empirical tight-binding calculations}

\author{Yoshiyuki Ohtsubo}
\email{y_oh@fbs.osaka-u.ac.jp}
\affiliation{Graduate School of Frontier Biosciences, Osaka University, Suita 565-0871, Japan}
\affiliation{Department of Physics, Graduate School of Sciences, Osaka University, Toyonaka 560-0043, Japan}
\author{Shin-ichi Kimura}
\email{kimura@fbs.osaka-u.ac.jp}
\affiliation{Graduate School of Frontier Biosciences, Osaka University, Suita 565-0871, Japan}
\affiliation{Department of Physics, Graduate School of Sciences, Osaka University, Toyonaka 560-0043, Japan}

\date{\today}

\begin{abstract} 
The topological order of single-crystal Bi and its surface states on the (111) surface are studied in detail based on empirical tight-binding (TB) calculations. New TB parameters are presented that are used to calculate the surface states of semi-infinite single-crystal Bi(111), which agree with the experimental angle-resolved photoelectron spectroscopy results. The influence of the crystal lattice distortion is surveyed and a topological phase transition is found that is driven by in-plane expansion. In contrast with the semi-infinite system, the surface-state dispersions on finite-thickness slabs are non-trivial irrespective of the bulk topological order. The role of the interaction between the top and bottom surfaces in the slab is systematically studied, and it is revealed that a very thick slab is required to properly obtain the bulk topological order of Bi from the (111) surface state: above 150 biatomic layers in this case.
\end{abstract}

\maketitle

\section{Introduction}
Topological materials classified by unconventional parity eigenvalues of three-dimensional (bulk) bands is one of the topics of high interest in solid state physics in this decade.
The insulators classified in the non-trivial (topological) group are called topological insulators (TIs), and hold metallic and spin-polarized surface states that continuously disperse between the bulk valence band maximum and the conduction band minimum \cite{Fu07, Hasan10, Qi11}.
Because these topological surface states are spin-polarized and robust against any perturbations that do not change the parities of the entire bulk band structure, these states are regarded as a promising element for future spintronic devices \cite{Manchon15}.

In the early stages of TI research, first-principles calculations based on density functional theory (DFT) achieved great success in predicting the topology of many materials and in proposing new TI candidates \cite{Zhang09, Yan10, Lin10, Eremeev10, Zhang13}. Most of these predictions were soon proven experimentally and there was excellent agreement between the predicted topological surface states and the observations \cite{Xia09, Hsieh09, Kuroda10}.
However, despite these great successes, there remains an open question on the topological order of the very simple material of single-crystal Bi.
Bismuth  is widely used as a component of TIs, such as Bi$_2$Se$_3$ \cite{Zhang09, Xia09} and TlBiSe$_2$ \cite{Yan10, Lin10, Eremeev10, Kuroda10}, because it is the heaviest non-radioactive element that also possesses a strong spin-orbit interaction (SOI). This is critical because the SOI plays an important role to realize the unconventional parity eigenvalues.
The topological order of single-crystal Bi has also been extensively studied together with its relative alloy Bi$_{1-x}$Sb$_x$, which is the first material experimentally detected as three-dimensional TI \cite{Hsieh08, Hsieh092}.
According to the DFT \cite{Fu07, Hsieh08} as well as empirical tight-binding (TB) \cite{Liu95, Teo08} calculations, the topological order of Bi is trivial, and alloying with Sb causes the topological phase transition to a TI to occur at $x \approx$ 0.04 .
However, unlike most of the other cases, this prediction is not consistent with the experimental results.

Figures 1(a) and 1(b) show the experimental surface-band dispersion of Bi(111) obtained by angle-resolved photoelectron spectroscopy (ARPES) \cite{Ohtsubo13}.
In these dispersions, the two spin-split surface bands $S1$ and $S2$ merge into the same projected bulk valence bands (BVBs) at $\bar{\Gamma}$.
However, at $\bar{M}$, these spin-split surface bands separately merge into two different projected bulk bands, with $S1$ merging into the bulk conduction band (BCB), while $S2$ merges into the BVB.
The overall dispersion is also schematically depicted in Fig. 1(d) .
Based on this surface-state dispersion, $S1$ continuously connects the projected BVB and BCB between two time-reversal-invariant momenta (TRIM), which is the behavior  that is expected for topologically non-trivial materials.
Indeed, a recent ARPES report on Bi$_{(1-x)}$Sb$_x$ ($x \sim$ 0.1) has demonstrated an almost identical surface-state dispersion \cite{Benia15}, with the sole difference being that the projected BCB is above the Fermi level at $\bar{M}$ in Bi$_{(1-x)}$Sb$_x$.
It should be noted that the topological classification of the band structure is not only valid for semiconductors, but also for semimetals whose finite projected bulk band gap opens in any $k_{\parallel}$ in the surface plane.
Interestingly, the surface-band dispersions observed by ARPES are qualitatively the same in various experiments on single-crystal Bi(111) \cite{Ast03} as well as thin films possessing a few tens of biatomic layers (BL) grown on Si(111) \cite{Hirahara06, Benia15}. Specifically, all the observations show that $S1$ merges into the BCB while $S2$ merges into the BVB at $\bar{M}$.

\begin{figure*}
\includegraphics[width=150mm]{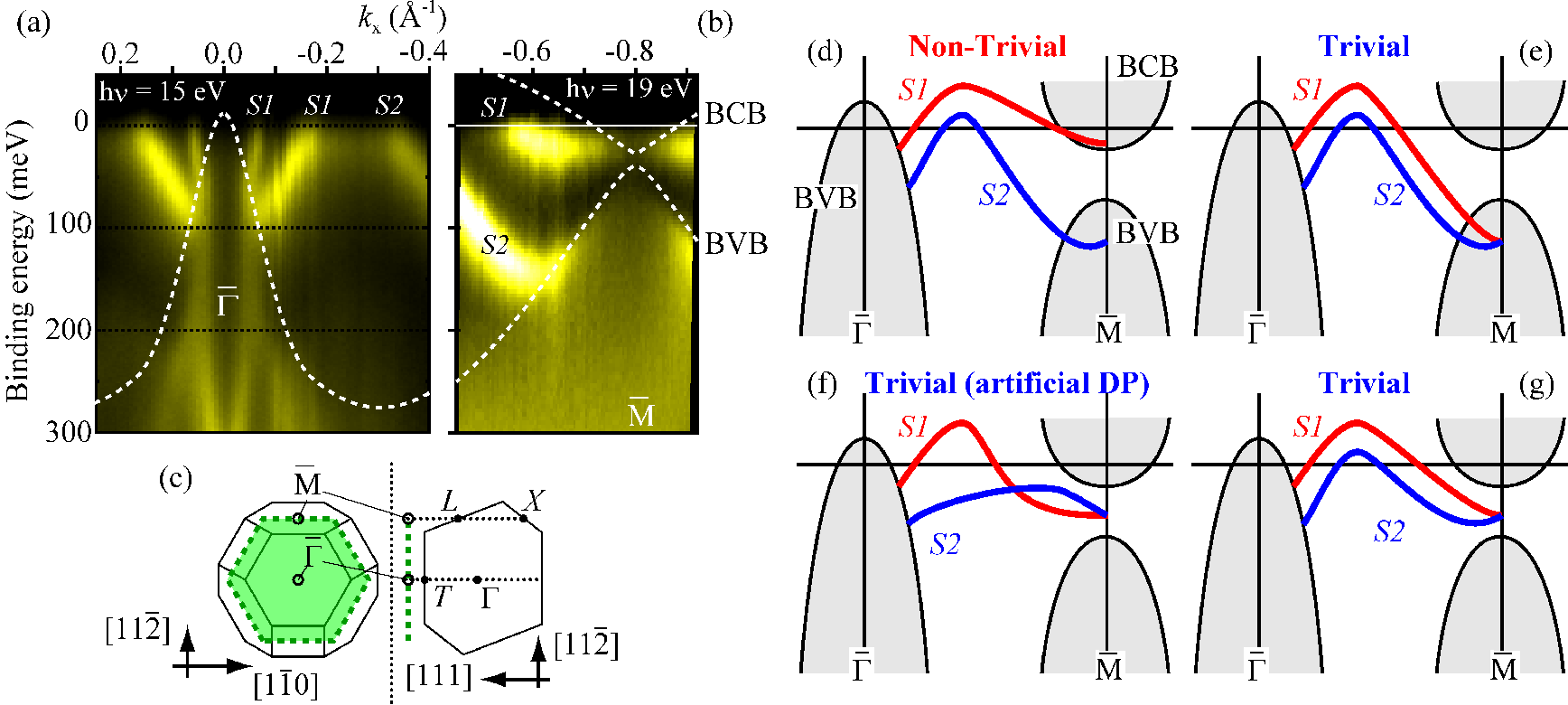}
\caption{\label{fig1} (a,b) Angle-resolved photoelectron spectroscopy intensity plots along $\bar{\Gamma}$-- $\bar{M}$ taken at 7.5 K. Dashed lines indicate the edge of projected bulk bands. These figures are reproduced from Ref. \onlinecite{Ohtsubo13}.
(c) Schematic drawing of the three-dimensional Brillouin zone (solid line) of the Bi single crystal and its projection onto the (111) surface Brillouin zone (dashed line).
(d-g) Schematic drawings of the surface-state dispersion and projected bulk bands along $\bar{\Gamma}$-- $\bar{M}$, obtained by these various methods:
(d,e) Based on local density approximation with slab geometry including (c) and excluding (d) the interaction between the top and bottom surface.
(f) Based on empirical tight-binding (TB) parameters and transfer-matrix method.
(g) The same as (e) but using different TB parameters.
}
\end{figure*}

The surface-state dispersions simulated by theoretical calculations depend upon the computational methods, however.
The DFT calculations based on the local density approximation (LDA) with SOI, using slab geometry to mimic the crystal surface, obtain the surface bands depicted in Fig. 1(e) \cite{Koroteev04, Hirahara06}.
In this case, both $S1$ and $S2$ merge into the BVB at $\bar{M}$, and hence the surface band dispersion is trivial.
This is consistent with the theoretical prediction mentioned above, but disagrees with the ARPES experiments.
The other major method used to calculate the surface state is the so-called transfer-matrix (TM) method with the empirical TB model for bulk electronic states \cite{Lee81, Teo08}.
Based on this method, however, the calculated surface state exhibits an additional crossing between $\bar{\Gamma}$ and $\bar{M}$, as shown in Fig. 1(f).
This unexpected crossing can be understood as the influence arising from an incorrect mirror Chern number via the empirical TB parameters \cite{Teo08}.
Further, another TM calculation based on a different set of TB parameters \cite{Hirahara12} has resulted in surface bands without this unrealistic surface-state crossing, and is shown in Fig. 1(g).
In this case, the Dirac point lies in the projected bulk band gap at $\bar{M}$.
The result in Fig. 1(g) is also a topologically trivial surface-band dispersion because no surface band continuously connects the BVB and BCB.

The main reason for the difficulty in calculating the electronic structure of single-crystal Bi is that its bandgap is very small. The bandgap for single-crystal Bi is $\sim$15 meV at $L$ in the bulk Brillouin zone, corresponding to $\bar{M}$ on the surface Brillouin zone as shown in Fig. 1(c).
One of the well-known weaknesses of DFT is its inability to estimate the accurate size of the bandgap.
Actually, LDA overestimates the size of the bandgap at $L$ \cite{Koroteev04}.
A recent study based on the quasiparticle self-consistent {\it GW} method including SOI has improved the size of the bandgap \cite{Aguilera15}.
Even using such state-of-the-art computational methods, however, the topological order of single-crystal Bi is still calculated to be trivial, which disagrees with the experimental results.
In the TB calculation, the size of the bandgap agrees with the experiments because the TB parameters are empirically tuned to reproduce these experimental results, although the topological order was also calculated to be trivial.

The tiny bulk bandgap at $L$ leads to ``flagile'' topological phase of single-crystal Bi, because various perturbation from strain, ultrathin film thickness and so on can invert the energetical order of the bulk bands at $L$.
Actually, a DFT calculation taking the inter-surface interaction in the slab into account showed non-trivial surface-band dispersion as shown in Fig. 1(d) \cite{Hirahara06}.
Recently, a TB calculation using the slab model also showed the surface states which disperses from BVB at $\bar{\Gamma}$ to BCB at $\bar{M}$ continuously \cite{Saito16}.
Since the bulk topological order based on DFT and TB with known parameter set is trivial, this result suggests the topological phase transition driven by ultrathin film thickness of Bi.
However, the magnitude of such finite-size effect, in other words, how thick the slab should be in order to calculate the surface states of Bi obeying the bulk topological order, has not been studied yet.
Structural strain is also claimed as a source of topological phase transition.
It is claimed that the in-plane structural strain in Bi(111) ultrathin film causes the topological phase transition from trivial to non-trivial phase \cite{Hirahara12, Yao16}.
A recent theoretical study supports this results \cite{Aguilera15}.
However, there is still a discontinuity to the bulk case without strain: based on the ARPES experiments, Bi without strain should be topologically non-trivial and hence it is not clear where the topological phase transition takes place with lattice distortion.

In this work, we present the new set of TB parameters for calculating the surface states of single-crystal Bi(111), which agrees with the experimental ARPES results.
Two sets of parameters were obtained that generate topologically trivial and non-trivial surface states for Bi, where neither exhibit any artificial crossing such as that generated previously (Fig. 1(e)) \cite{Liu95, Teo08}.
Based on the new TB parameters, the surface-state calculations were performed by means of both the TM method and slab geometry.
In both cases, the calculated surface states agree well with previous experimental results, except for in the proximity of $\bar{M}$.
Around $\bar{M}$, the surface-state dispersion changes depending upon the topological order of the bulk bands, where only the topologically non-trivial case agrees with the experiments.
The influence of crystal lattice distortion was surveyed and a topological phase transition was found that was driven by in-plane expansion for the non-trivial bulk bands, which was opposite to the distortion suggested in a previous report \cite{Hirahara12}.
In contrast with the semi-infinite crystal, the slab calculation generated non-trivial surface-state dispersions irrespective of the bulk topological order, as has been the case for previous DFT calculations using slab geometries.
The role of the interaction between the top and bottom surfaces in the slab was systematically studied, and it was found that a very thick slab is required to obtain the bulk topological order of Bi from the (111) surface states properly. Specifically, the slab must be greater than 150 BL in our model.

\section{Computational methods}

\begin{figure}
\includegraphics[width=80mm]{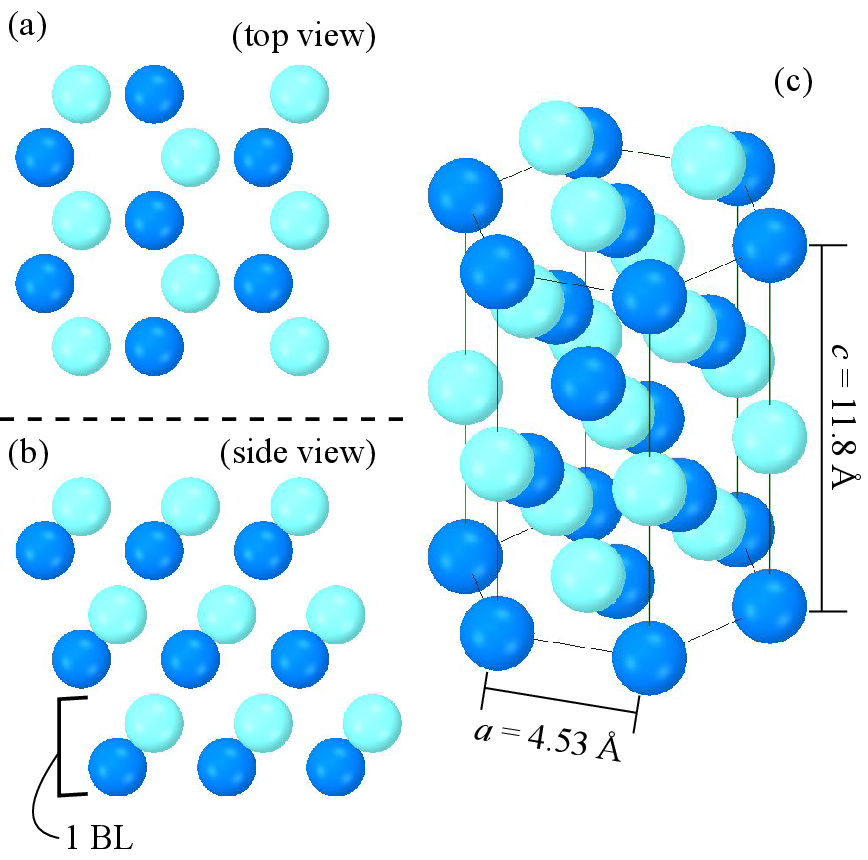}
\caption{\label{fig2} Crystal structure of Bi. The lower contrast circles (light blue) represent the second atoms in the primitive unit cell (see text) in the (a) top and (b) side view of Bi(111).
}
\end{figure}

\subsection{Tight-binding parameters for bulk states}
The main framework used to calculate the bulk electronic structure was the same as that in Ref. \onlinecite{Liu95} , and is briefly explained herein.
Single-crystal Bi has an A17 rhombohedral lattice, but is also characterized by hexagonal lattice parameters, {\it a} and {\it c}, together with an additional parameter $\mu$ (see Fig. 2).
The primitive (rhombohedral) unit cell contains two atoms and the relative position of the second atom is (0, 0, 2$\mu c$), where $\mu$ = 0.2341.
The hopping parameters of the $sp^3$ orbitals between first-, second- and third-nearest-neighbor  atoms were taken, and the SOI was included via the spin-orbit coupling parameter $\lambda$ = 1.5 eV.
The resulting (16$\times$16) matrix is shown in the appendix of Ref. \onlinecite{Liu95} .

Next, the TB parameters were modified so that the calculated surface-state dispersion between $\bar{\Gamma}$ and $\bar{M}$ could be reproduced without any artificial crossing such as was obtained in Ref. \onlinecite{Teo08} .
In addition, the new parameters were tuned to maintain the energies of the electron/hole pockets and the size of the bandgap at $L$ at nearly the same value as the original parameter, which agrees with the experimental values.
Table I represents two sets of the new TB parameters, labeled as TBP-1 and TBP-2 in the following, obtained by the methods above.
Table II represents the parity invariants at each TRIM in the Brillouin zone of the bulk Bi crystal, calculated with the parameters given in Table I.
As shown the $\nu_0$ values in Table II, the major difference between TBP-1 and TBP-2 is the difference of  topological order: trivial for TBP-1 but non-trivial for TBP-2.

\begin{table}
\caption{\label{tab:one} The new sets of tight-binding (TB) parameters for single-crystal Bi tuned so that the they could reproduce the surface states which agree with the experimental results. The definitions of each parameter are the same as in Ref. \onlinecite{Liu95}.}
\begin{ruledtabular}
\begin{tabular}{cccc}
Parameter & Ref. \onlinecite{Liu95} & TBP-1 & TBP-2  \\
(eV)& (trivial) & (trivial) & (non-trivial) \\ 
\hline
$E_s$ & $-$ 10.906 & $-$10.906 & $-$10.710 \\
$E_p$ & $-$0.486 & $-$0.336 & $-$0.366 \\
$V_{ss\sigma}$ & $-$0.608 & $-$2.860 & $-$2.860 \\
$V_{sp\sigma}$ & 1.320 & 1.308 & 1.340 \\
$V_{pp\sigma}$ & 1.854 & 1.855 & 1.844 \\
$V_{pp\pi}$ & $-$0.600 & $-$0.600 & $-$0.600 \\
$V'_{ss\sigma}$ & $-$0.384 & $-$0.384 & $-$0.384 \\
$V'_{sp\sigma}$ & 0.433 & $-$0.100 & $-$0.050 \\
$V'_{pp\sigma}$ & 1.396 & 1.396 & 1.382 \\
$V'_{pp\pi}$ & $-$0.344 & $-$0.344 & $-$0.344 \\
$V''_{ss\sigma}$ & 0 & 0 & 0 \\
$V''_{sp\sigma}$ & 0 & 0 & 0.300 \\
$V''_{pp\sigma}$ & 0.156 & 0.156 & 0.156 \\
$V''_{pp\pi}$ & 0 & $-$0.050 & $-$0.040 \\
\end{tabular}
\end{ruledtabular}
\end{table}

\begin{table}
\caption{\label{tab:two} Parity invariants ($\delta$) at each time-reversal-invariant momentum (TRIM; $\Gamma , L, X, T$ ) and the $Z_2$ topological invariants ($\nu_0; \nu_1 \nu_2 \nu_3$) calculated with the tight-binding parameters shown in Table I.}
\begin{ruledtabular}
\begin{tabular}{cccccc}
 & $\delta$($\Gamma$) & $\delta$($L$) & $\delta$($X$) & $\delta$($T$) & ($\nu_0; \nu_1 \nu_2 \nu_3$) \\
\hline
Ref. \onlinecite{Liu95} & $-$ 1 & $-$1 & $-$1 & $-$1 & (0;000) \\
TBP-1 & $-$1 & $-$1 & $-$1 & $-$1 & (0;000) \\
TBP-2 & $-$1 & +1 & $-$1 & $-$1 & (1;111) \\
\end{tabular}
\end{ruledtabular}
\end{table}

\subsection{Transfer-matrix (TM) method}
The TM method is used to calculate the surface electronic states on a semi-infinite crystal from the bulk Hamiltonian and the transfer matrix $T(k_{\parallel}, E)$, as proposed in Ref. \onlinecite{Lee81} . In this TM, $k_{\parallel}$ is the in-plane wavevector and $E$ is the binding energy .
The procedure reported in a previous paper \cite{Teo08} was followed, as described below, but the bulk TB parameters were changed.

Because the Bi crystal can be regarded as a stack of BLs, as depicted in Fig. 2(b), the bulk electronic states of a semi-infinite Bi crystal can be written as
\begin{eqnarray}
\left( \begin{array}{c}
\phi_{n+1, 1} \\
\phi_{n+1, 2} \\
\end{array}
\right)
= T(k_{\parallel}, E)
\left( \begin{array}{c}
\phi_{n, 1} \\
\phi_{n, 2} \\
\end{array}
\right),
\end{eqnarray}
where $\phi_{n, a}$ is a basis of the states in the BL plane localized on the $a$ = 1, 2 monolayer of the $n$th BL.
In addition, each $\phi_{n, a}$ has eight components associated with the eight atomic orbitals.
The transfer matrix, $T(k_{\parallel}, E)$, is given by Equations (3.1) to (3.3) in Ref. \onlinecite{Teo08} , together with the appendix in Ref. \onlinecite{Liu95} .
Any bulk states are the eigenstates of the 16$\times$16 TM with unimodular eigenvalues.
For each $E$ in the projected bulk bandgap, $T(k_{\parallel}, E)$ has eight eigenvalues with moduli larger than 1, which correspond to the electronic states whose amplitude decays in the $-z$  direction.
With the boundary condition $\phi_{0,1}=0$, the surface states should also decay outside the crystal. These surface states are determined by forming an 8$\times$8 matrix $M(k_{\parallel}, E)$ composed of the eight components of $\phi_{0,1}$ for each of the eight decaying states.
The detailed procedure to generate $M(k_{\parallel}, E)$ is shown in Ref. \onlinecite{Lee81}.
Finally, the surface-state band dispersion ($E(k_{\parallel})$)  is determined by solving det[$M(k_{\parallel}, E)$] =0.

\subsection{Finite slab calculation}
Slab geometry is widely used for DFT calculations of surface electronic structures, such as used in Refs. \onlinecite{Hirahara06, Koroteev04, Hirahara12, Saito16} .
This model can mimic the surface without breaking the three-dimensional periodicity and can therefore be easily applied toward various surface systems.
However, sometimes this model generates artificial states owing to the interaction between the top and bottom surfaces.
In this work, we followed the method reported in the recent paper \cite{Saito16}, as is described below.
Varying from the previous work, however, we used a new set of bulk TB parameters and assumed no surface hopping term.

The total Hamiltonian of the slab $H$ is represented as
\begin{eqnarray}
H=
\left( \begin{array}{cccccc}
H_{11} & H_{12}^{(1)} & 0 & 0 & \cdots & 0\\
H_{21}^{(1)} & H_{11} & H_{21}^{(2)} & 0 & \cdots & 0\\
0 & H_{12}^{(2)} & H_{11} & H_{12}^{(1)} & \cdots  & 0\\
\vdots &  & \ddots & \ddots & \ddots &  \\
0 & \cdots & 0  & H_{12}^{(2)} & H_{11} & H_{12}^{(1)} \\
0 & \cdots & 0  & 0 & H_{21}^{(1)} & H_{11} \\
\end{array}
\right),
\end{eqnarray}
where $H_{11}$ is the hopping term for inter-monatomic-layer hopping (i.e., in the same monatomic layer) and $H_{12}^{(1)}$ ($H_{21}^{(2)}$) are the intra-BL (inter-BL) hopping terms.
The $H_{12}^{(1)}$, $H_{21}^{(2)}$, and $H_{11}$ terms are given in the bulk TB Hamiltonian \cite{Liu95} as the first-, second- and third-nearest-neighbor hopping terms, respectively.
The size of the matrix is therefore 16$n\times$16$n$, where $n$ is the number of the BLs in the slab.

In the following, the obtained states were plotted with Gaussian broadening ($FWHM$ = 5 meV) to mimic the ARPES intensity plots produced with a typical instrumental energy resolution.
\section{Results and discussion}

\subsection{Surface states generated by TM method}
\begin{figure}
\includegraphics[width=80mm]{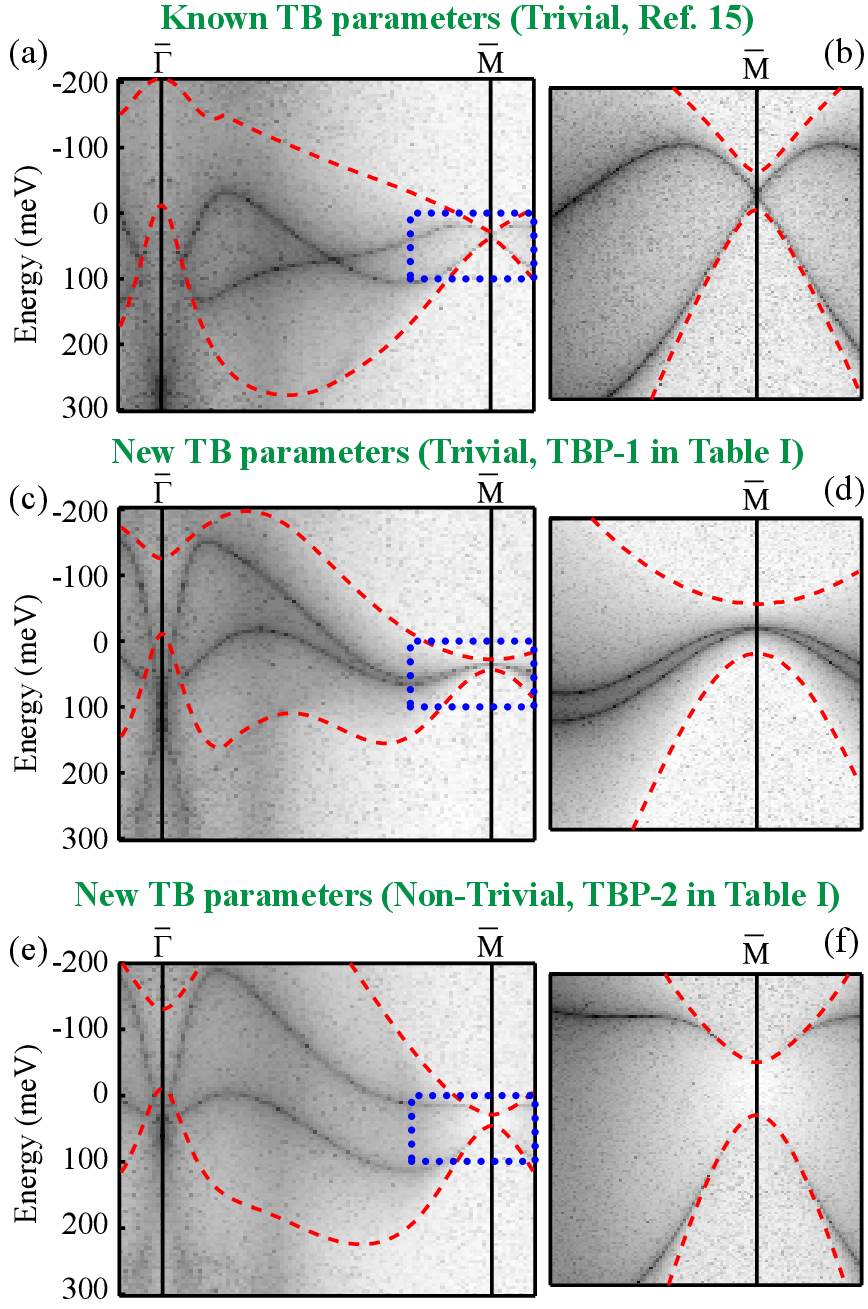} 
\caption{\label{fig3} Surface-state band dispersions calculated via the transfer-matrix method. 
The intensities $ln$(1/det[$M(k_{\parallel}, E)$]) were plotted so that the locations $(k_{\parallel}, E)$ where surface states lie are less  (i.e., darker in grayscale intensity plot) than the others.
Dashed lines represent the edges of projected bulk bands.
(a) Surface states calculated using the tight-binding (TB) parameters from Ref. \onlinecite{Liu95}.
(b) A close-up image around $\bar{M}$. The plotted area is indicated by a dotted square in (a).
(c,d) The same as (a) and (b), respectively, but calculated using the new TB parameters generating topologically trivial bulk bands (TBP-1 in Table I). 
(e,f) The same as (a) and (b), respectively, but calculated using the new TB parameters generating topologically non-trivial bulk bands (TBP-2 in Table I). 
}
\end{figure}

Figure 3 plots the surface states and the edge of the projected bulk bands generated using the TM method with three different sets of bulk TB parameters.
To plot the surface-state dispersion, we plotted $ln$(1/det[$M(k_{\parallel}, E)$]), so that the locations $(k_{\parallel}, E)$ where surface states lie possess much smaller intensity values (i.e., darker) than the others .

The surface bands calculated with the TB parameters given in Ref. \onlinecite{Liu95}  (Figs. 3(a) and 3(b)) agree with those given in the previous paper exhibiting an artificial crossing of the surface bands between $\bar{\Gamma}$ and $\bar{M}$ owing to an incorrect mirror Chern number \cite{Teo08}.
This artificial surface-band crossing disappears when the new TB parameters are used.
Using the parameters that generate topologically trivial bulk bands (TBP-1 in Table I), the surface-band dispersion qualitatively agrees with the ARPES results except for the $k$ region around $\bar{M}$, as shown in Figs. 3(c) and 3(d).
The two branches of the surface bands degenerate with each other at $\bar{M}$, and this surface-state dispersion is therefore topologically trivial, as is expected from the topological order of bulk bands.
This surface-state dispersion agrees with that reported in a previous paper \cite{Hirahara12}.
Around $\bar{\Gamma}$, the TB parameters generating non-trivial bulk bands (TBP-2 in Table I) does not significantly alter the surface-state dispersion from the trivial dispersion, as shown in Fig. 3(e).
Around $\bar{M}$,however, the surface-state dispersion is different from those calculated with the other TB parameters. Specifically, the upper branch merges into the BCB while the lower merges into the BVB, showing a good agreement with the experimental results \cite{Ast03, Hirahara06, Ohtsubo13, Benia15} (see Fig. 3(f)).

These good agreements of surface-state dispersions with the previous experimental and theoretical results except for the proximity of $\bar{M}$ implies that the TB calculation can no longer be the evidence of topological order of single-crystal Bi.
In order to judge such balancing two possibilities, one requires the experimental data. 
Based on the ARPES results, it should be topologically non-trivial, while ARPES does not provide any direct information about the parity eigenvalues of bulk bands at $L$.
The other, bulk-sensitive and accurate experimental method would be helpful to make a firm conclution on this controversial issue, topological order of single-crystal Bi.

\subsection{Topological phase transition driven by lattice distortion}

\begin{figure}
\includegraphics[width=80mm]{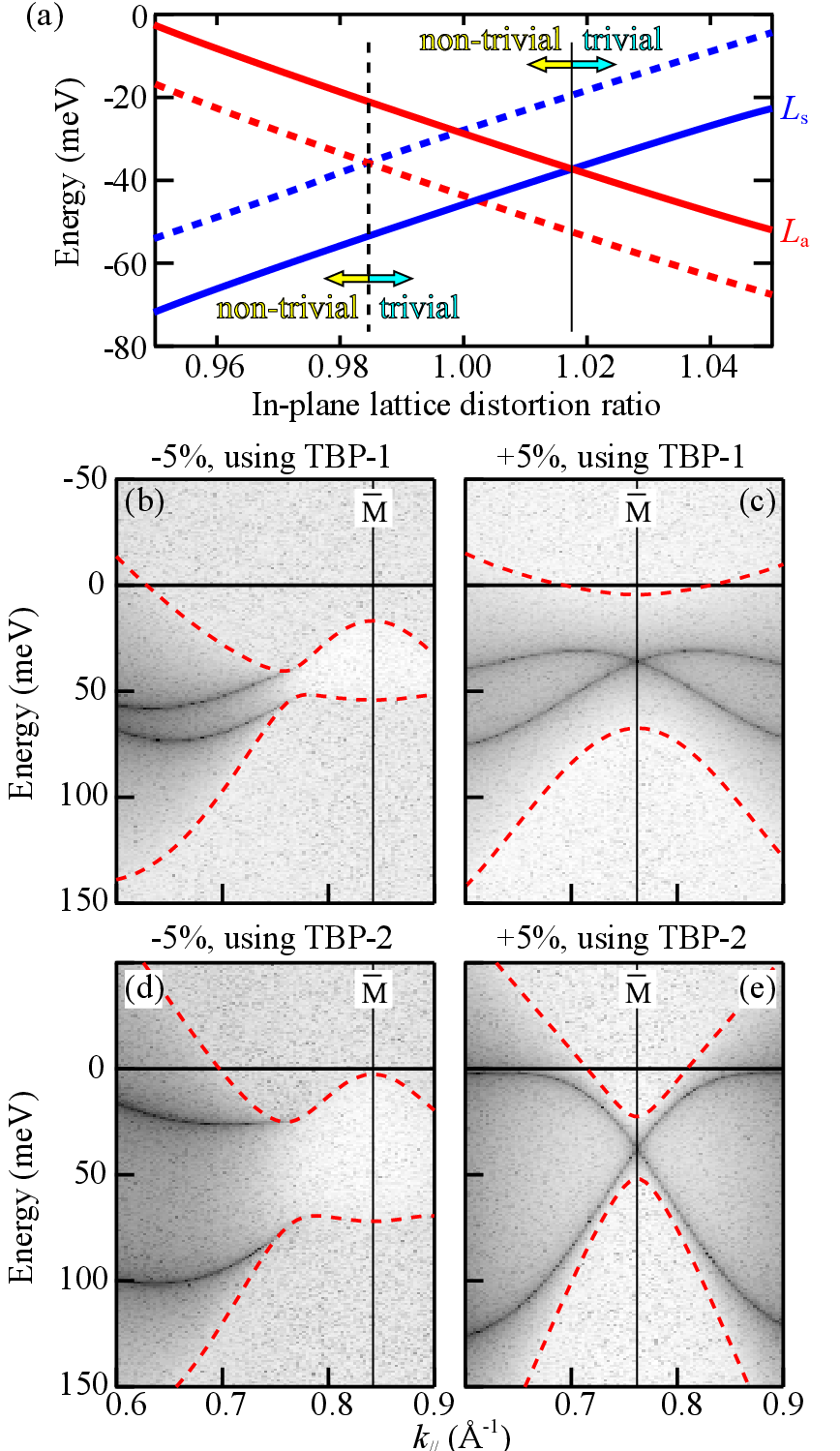} 
\caption{\label{fig4} (a) Band evolution at $L$. Solid (dashed) lines are the energies of bulk bands just above and below the bandgap at $L$ calculated using the tight-binding (TB) parameter set TBP-1 (TBP-2) in Table I that give topologically non-trivial (trivial) topological order .
Vertical lines indicate the position where the topological phase transition occurs, corresponding with each TB parameter set.
(b--e) Surface-state band dispersions calculated by the transfer-matrix method with in-plane lattice distortions of (b,d) $-$5 \% lattice strain and (c,e) +5 \% lattice expansion.
The dispersions are obtained using the TB parameter set (b,c) TBP-1 and (d,e) TBP-2, given in Table I.
}
\end{figure}

In order to examine the topological phase transition driven by structural distortion based on Refs. \onlinecite{Hirahara12, Aguilera15, Yao16}, we surveyed the topological order and surface-state dispersion around $\bar{M}$ using the TM calculation and the new TB parameters.

Figure 4(a) shows the bulk band evolution at $L$ with an in-plane lattice distortion inserted for calculations using the two TB parameter sets in Table I, TBP-1 and TBP-2. 
Irrespective of the topological order generated with zero lattice distortion, a topological phase transition occurs in both cases.
The only difference observed is whether the topological phase transition occurs when one uses a lattice strain (TBP-1, trivial at zero lattice distortion) or a lattice expansion (TBP-2, non-trivial at zero lattice distortion). 
Figures 4(b--e) show the surface-state dispersions around $\bar{M}$ obtained using the TM calculation as Fig. 3.
Note that the $\bar{M}$ position of each plot changes according to the in-plane lattice constant.
The plots in Figs. 4(b) and 4(d) show that the surface-state dispersion is topologically non-trivial with lattice strain ($-$5 \%) for both TB parameter sets.
In addition, with an in-plane lattice expansion (+5 \%), the surface states for both TB parameter sets exhibit a trivial dispersion, as seen in Figs. 4(c) and 4(e), that form a Kramers-degeneracy point at $\bar{M}$ in the projected bulk bandgap.
It should be noted that the non-trivial surface-state dispersion observed by ARPES experiments with the presence of in-plane lattice strain \cite{Hirahara12, Yao16} exhibits no conflict with the calculated results from both of the new TB parameters, TBP-1 and TBP-2.

Based on these results, we propose that to trace the bulk bandgap of single-crystal Bi with in-plane lattice distortion in order to conclude the topological order of Bi.
If the bandgap closes with the in-plane lattice expantion (strain), it would be the smoking-gun evidence of non-trivial (trivial) topological order.

\begin{figure}
\includegraphics[width=80mm]{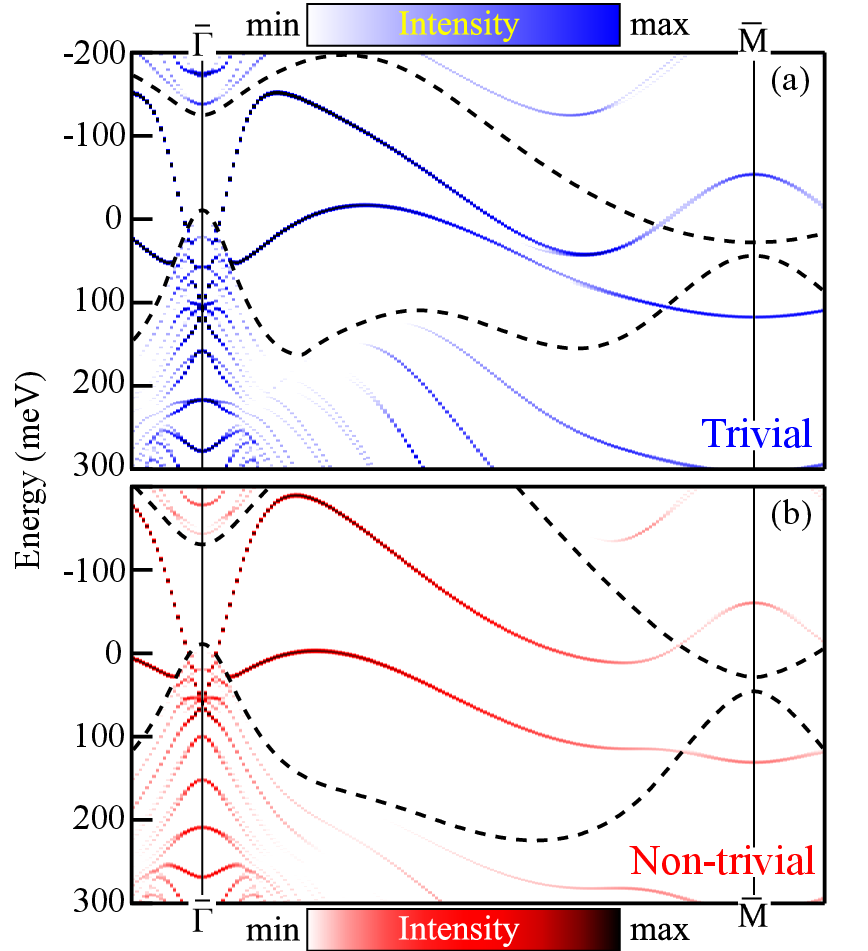}
\caption{\label{fig5} Band dispersions calculated for a 20 biatomic-layer (BL) slab using the new tight-binding parameters that generate the (a) trivial (TBP-1 in Table I) and (b) non-trivial (TBP-2 in Table I) bulk topological order.
Dashed lines represent the edge of the projected bulk bands.
Intensities are obtained from the eigenfunction amplitude localized in the topmost surface BL.
}
\end{figure}

\subsection{Surface states on a finite slab}

Figure 5 shows the electronic structure calculated with the slab geometry, where the slab thickness is 20 BL and the dashed lines represent the edge of the projected bulk bands.
The surface-state bands dispersing in the projected bulk bandgap are obtained together with the discrete quantum well (QW) states corresponding to the bulk bands in the projected bulk-band region.
Around $\bar{\Gamma}$, the surface-state dispersions are almost identical to those calculated by the TM method (see Figs. 3(c) and 3(e)).
Note that no additional surface hopping terms are needed to obtain these surface bands, which is in contrast with a previous study \cite{Saito16}.
The fact that no additional hopping terms are needed is possibly owing to the different TB parameter set used in this study;
however, the surface-state dispersions are quite different around $\bar{M}$.
Even when using the TB parameter that generates a trivial topology of bulk bands (TBP-1 in Table I), the surface-state dispersion obtained by the slab geometry suggests a non-trivial topological order wherein the upper branch merges into the BCB while the lower merges into the BVB.
Such behavior  is the same as that reported in previous studies \cite{Hirahara06, Saito16}, wherein 
it has been explained as the influence of the interaction between the top and bottom surfaces owing to the finite slab thickness.
Even when using the trivial TB parameter set (TBP-1 in Table I), the inter-surface interaction would re-invert  the bulk bands at $L$, where the bulk bandgap is the smallest, and thus cause the topological phase transition.
Similar topological phase transitions driven by the film thickness have been observed in the QWs of HgTe \cite{Konig07, Chu11}.

\begin{figure}
\includegraphics[width=80mm]{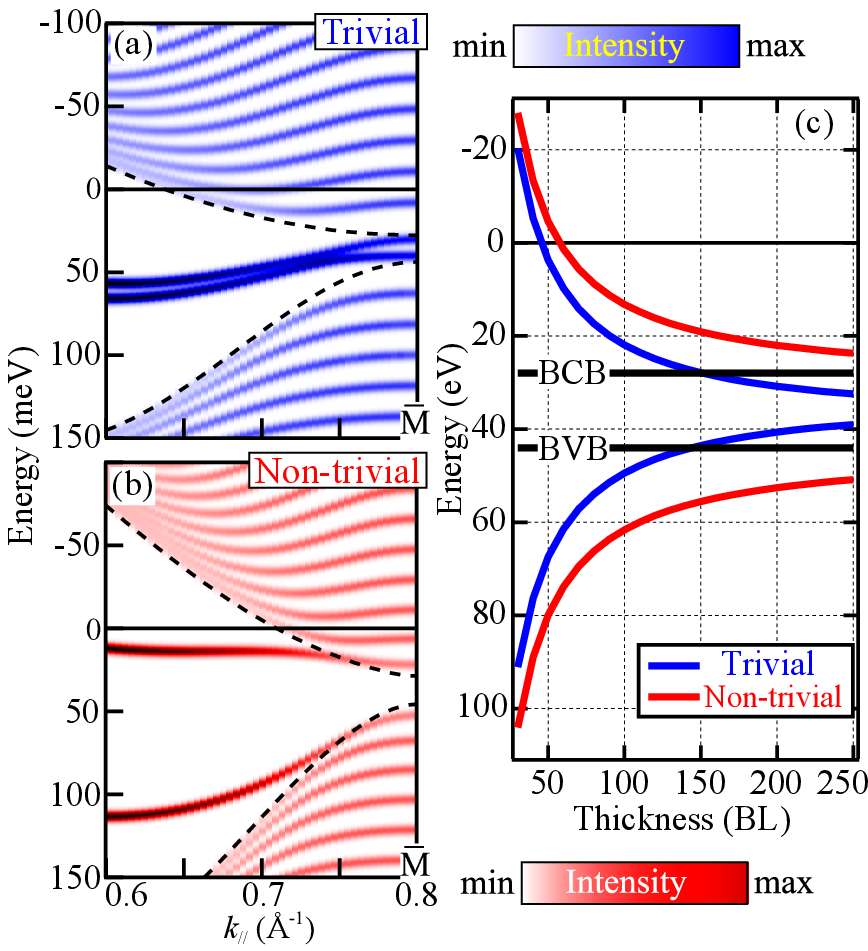}
\caption{\label{fig6} The electronic structure around $\bar{M}$ in a 200 biatomic-layer Bi slab using the new tight-binding parameters that give (a) trivial (TBP-1 in Table I) and (b) non-trivial (TBP-2 in Table I) bulk topological order  (cf. Figs. 5(a) and 5(b), respectively).
(c) Energy evolutions of the quantum-well-like states at $\bar{M}$ connected to the surface-state bands in the projected bulk band gap.
Thick horizontal lines indicate the energies of the bottom of the projected bulk conduction band and the top of the projected bulk valence band.
}
\end{figure}

To explore the finite-thickness effect in more detail, we calculated the electronic states around $\bar{M}$ at different thicknesses.
Figures 6(a) and 6(b) plot the electronic structure around $\bar{M}$ in a 200 BL slab calculated with the TB parameters TBP-1 and TBP-2, respectively, which are given in Talbe I.
The number of the QW states in these plots are much larger than those of the 20-BL slab (cf. Figure 5), reflecting the increased slab thickness.
In the proximity of $\bar{M}$, however, one can find the difference between Figs. 6(a) and 6(b).
The two surface-state branches dispersing out of the projected bulk bands in Fig. 6(b) enter the projected bulk bands and become QW states.
However, in Fig. 6(a), these branches do not enter the projected bulk bands but remain in the bulk bandgap at $\bar{M}$, as is the case for the semi-infinite TM calculation (cf. Fig. 3(d)).
Figure 6(c) plots the energy evolution of these two surface states at $\bar{M}$, which are connected to the surface states away from $\bar{M}$, together with the energy positions of the projected BVB and BCB.
As shown in Fig. 6(c), the surface states calculated with TBP-1 (trivial bulk bands) disperse out of the projected bulk bands at a slab thickness greater than $\sim$150 BL.
In contrast, with TBP-2 (topologically non-trivial bulk bands), the surface states never appear out of the projected bulk bands at $\bar{M}$.
This result suggests that a much thicker slab is required to accurately measure the topological order of single-crystal Bi than has been used in previous studies \cite{Hirahara12, Yao16}. With a slab that is insufficiently thick, the surface-state dispersion always exhibits the topologically non-trivial behavior  irrespective of the topological order of the bulk Bi.

Recently, two groups have reported ARPES experimental results of the Bi(111) surface states using a thickness above 100 BL \cite{Hirahara15, Ito16}.
In both cases, the surface-state bands disperse as that seen in Fig. 6(b), and hence these experimental results indicate the non-trivial topological order of Bi, which is in good agreement with the bulk single-crystal case \cite{Ohtsubo13}.

\section{Summary}
In summary, new TB parameters are presented with which to calculate the surface states of single-crystal Bi(111) that agree with the experimental ARPES results.
Two sets of TB parameters were obtained that make Bi topologically trivial or non-trivial, wherein neither parameter set exhibits any artificial crossings such as those generated by previous TB parameters (Fig. 1(e)) \cite{Liu95, Teo08}.
Based on the new TB parameters, surface-state calculations were performed using both the transfer-matrix method and slab geometry.
In both cases, the calculated surface states agree well with the previous experimental results, except for in the proximity of $\bar{M}$.
Around $\bar{M}$, the surface-state dispersion changes depending upon the topological order of the bulk bands, wherein only the topologically non-trivial case agrees with the experiments.
We surveyed the influence of crystal lattice distortions and found a topological phase transition driven by in-plane expansion for the non-trivial bulk bands, which is opposite to the distortion found in a previous report \cite{Hirahara12}.
In contrast with a semi-infinite crystal, the slab calculation generated non-trivial surface-state dispersions irrespective of the bulk topological order, as is the case suggested in previous DFT calculations using slab geometries.
The role played by the interaction between the top and bottom surfaces in the slab was systematically studied and it was revealed that a very thick slab (i.e., greater than 150 BL in our case) is required to accurately obtain the bulk topological order of Bi from the (111) surface states.
These detailed calculations of the electronic structure and topological order of the simple, well-known material of single-crystal Bi will be helpful in the further research of topological materials.

\section*{Acknowledgement}

This work was supported by the Japan Society for the Promotion of Science Grants-in-Aid (B) and Scientific Research Activity Start-up (Nos. JP26887024 and JP15H03676, respectively) and The Murata Science Foundation.



\begin{thebibliography}{99} 
\bibitem{Fu07}
L. Fu and C. L. Kane, Phys. Rev. B {\bf 76}, 045302 (2007).
\bibitem{Hasan10}
M. Z. Hasan and C. L. Kane, Rev. Mod. Phys. {\bf 82}, 3045 (2010).
\bibitem{Qi11}
X.-L. Qi and S.-C. Zhang, Rev. Mod. Phys. {\bf 83}, 1057 (2011).
\bibitem{Manchon15}
A. Manchon, H. C. Koo, J. Nitta, S. M. Frolov and R. A. Duine, Nature Mat. {\bf 14}, 871 (2015).
\bibitem{Zhang09}
H. Zhang, C.-X. Liu, X.-L. Qi, X. Dai, Z. Fang and S.-C. Zhang, Nature Phys. {\bf 5}, 438 (2009).
\bibitem{Zhang13}
H. Zhang and S.-C. Zhang, Phys. Status Solidi RRL {\bf 7}, 72 (2013).
\bibitem{Yan10}
B. Yan, C.-X. Liu, H.-J. Zhang, C.-Y. Yam, X.-L. Qi, T. Frauenheim and S.-C. Zhang, Europhys. Lett. {\bf 90}, 37002 (2010).
\bibitem{Lin10}
H. Lin, R. S. Markiewicz, L. A. Wray, L. Fu, M. Z. Hasan and A. Bansil, Phys. Rev. Lett. {\bf 105}, 036404 (2010).
\bibitem{Eremeev10}
S. V. Eremeev, Y. M. Koroteev and E. V. Chulkov, JETP Lett. {\bf 91}, 594 (2010).
\bibitem{Xia09}
Y. Xia, D. Qian, D. Hsieh, L. Wray, A. Pal, H. Lin, A. Bansil, D. Grauer, Y. S. Hor, R. J. Cava and M. Z. Hasan , Nature Phys. {\bf 5}, 398 (2009)
\bibitem{Hsieh09}
D. Hsieh, Y. Xia, D. Qian, L. Wray, J. H. Dil, F. Meier, J. Osterwalder, L. Patthey, J. G. Checkelsky, N. P. Ong, A. V. Fedorov, H. Lin, A. Bansil, D. Grauer, Y. S. Hor, R. J. Cava and M. Z. Hasan , Nature {\bf 460}, 1101 (2009).
\bibitem{Kuroda10}
K. Kuroda, M. Ye, A. Kimura, S. V. Eremeev, E. E. Krasovskii, E. V. Chulkov, Y. Ueda, K. Miyamoto, T. Okuda, K. Shimada, H. Namatame and M. Taniguchi , Phys. Rev. Lett. {\bf 105}, 146801 (2010).
\bibitem{Hsieh08}
D. Hsieh, D. Qian, L. Wray, Y. Xia, Y. S. Hor, R. J. Cava and M. Z. Hasan, Nature {\bf 452}, 970 (2008).
\bibitem{Hsieh092}
D. Hsieh, Y. Xia, L. Wray, D. Qian, A. Pal, J. H. Dil, J. Osterwalder, F. Meier, G. Bihlmayer, C. L. Kane, Y. S. Hor, R. J. Cava and M. Z. Hasan , Science {\bf 323}, 919 (2009).
\bibitem{Liu95}
Y. Liu and R. E. Allen, Phys. Rev. B {\bf 52}, 1566 (1995).
\bibitem{Teo08}
J. C. Y. Teo, L. Fu and C. L. Kane, Phys. Rev. B {\bf 78}, 045426 (2008).
\bibitem{Ohtsubo13}
Y. Ohtsubo, L. Perfetti, M. O. Goerbig, P. Le F\`evre, F. Bertran and A. Taleb-Ibrahimi, New J. Phys. {\bf 15}, 033041 (2013).
\bibitem{Benia15}
H. M. Benia, C. Stra\ss er, K. Kern and C. R. Ast, Phys. Rev. B {\bf 91}, 161406(R) (2015).
\bibitem{Ast03}
C. R. Ast and H. H\"{o}chst, Phys. Rev. B {\bf 67}, 113102 (2003).
\bibitem{Hirahara06}
T. Hirahara, T. Nagao, I. Matsuda, G. Bihlmayer, E. V. Chulkov, Y. M. Koroteev, P. M. Echenique, M. Saito and S. Hasegawa, Phys. Rev. Lett. {\bf 97}, 146803 (2006).
\bibitem{Koroteev04}
Y. M. Koroteev, G. Bihlmayer, J. E. Gayone, E. V. Chulkov, S. Bl\"{u}gel, P. M. Echenique and Ph. Hofmann, Phys. Rev. Lett. {\bf 93}, 046403 (2004).
\bibitem{Lee81}
D. H. Lee and J. D. Joannopoulos, Phys. Rev. B {\bf 23}, 4988 (1981).
\bibitem{Hirahara12}
T. Hirahara, N. Fukui, T. Shirasawa, M. Yamada, M. Aitani, H. Miyazaki, M. Matsunami, S. Kimura, T. Takahashi, S. Hasegawa and K. Kobayashi , Phys. Rev. Lett. {\bf 109}, 227401 (2012).
\bibitem{Aguilera15}
I. Aguilera, C. Friedrich and S. Bl\"{u}gel, Phys. Rev. B {\bf 91}, 125129 (2015).
\bibitem{Saito16}
K. Saito, H. Sawahata, T. Komine and T. Aono, Phys. Rev. B {\bf 93}, 041301(R) (2016).
\bibitem{Yao16}
M. -Y. Yao, F. Zhu, C. Q. Han, D. D. Guan, C. Liu, D. Qian and J. Jia, Sci. Rep. {\bf 6}, 21236 (2016).

\bibitem{Konig07}
M. K\"{o}nig, S. Wiedmann, C. Br\"{u}ne, A. Roth, H. Buhmann, L. W. Molenkamp, X.-L. Qi and S.-C. Zhang, Science {\bf 318}, 766 (2007).
\bibitem{Chu11}
R.-L. Chu, W.-Y. Shan, J. Lu and S.-Q. Shen, Phys. Rev. B {\bf 83}, 075110 (2011)

\bibitem{Hirahara15}
T. Hirahara, T. Shirai, T. Hajiri, M. Matsunami, K. Tanaka, S. Kimura, S. Hasegawa and K. Kobayashi, Phys. Rev. Lett. {\bf 115}, 106803 (2015).
\bibitem{Ito16}
S. Ito, B. Feng, M. Arita, A. Takayama, T. Someya, W.-C. Chen, C.-M. Cheng, C.-H. Lin, S. Yamamoto, T. Iimori, H. Namatame, M. Taniguchi, S.-J. Tang, F. Komori, K. Kobayashi and I. Matsuda , arXiv: 1605.03531v1 (2016).

\end{thebibliography}
\end{document}